\documentclass[sigconf]{acmart}

\AtBeginDocument{%
  }

\newcommand{\ourbenchmark}{\texttt{\textsc{MOSAIC}}\xspace}

\usepackage{todonotes}
\usepackage{makecell}
\usepackage{subcaption}
\usepackage{xspace}
\usepackage{enumitem}
\usepackage{multirow}
\usepackage{adjustbox}
\usepackage{balance}
\usepackage{mdframed}

\usepackage{array}

\usepackage{booktabs}
\usepackage{pifont}
\usepackage{wasysym}
\newcommand{\cmark}{\ding{51}} 
\newcommand{\xmark}{\ding{55}} 
\newcommand{\pmark}{\LEFTcircle} 

\newcolumntype{C}{>{\centering\arraybackslash}p{2cm}}

\newcounter{boxcounter}

\copyrightyear{2026}
\acmYear{2026}
\setcopyright{cc}
\setcctype{by}
\acmConference[CIKM '26]{Proceedings of the 35th ACM International Conference on Information and Knowledge Management}{November 07--11, 2026}{Rome, Italy}
\acmBooktitle{Proceedings of the 35th ACM International Conference on Information and Knowledge Management (CIKM '26), November 07--11, 2026, Rome, Italy}
\acmDOI{10.1145/3799682.3840183}
\acmISBN{979-8-4007-2539-5/2026/11}

\begin{document}

\title{\ourbenchmark: Unveiling the Moral, Social and Individual Dimensions of Large Language Models}

\author{Erica Coppolillo}
\affiliation{%
  \institution{ICAR-CNR}
  \city{Rende}
  \country{Italy}}
\email{ericacoppolillo@cnr.it}

\author{Emilio Ferrara}
\affiliation{%
  \institution{University of Southern California}
  \city{Los Angeles, California}
  \country{USA}}
\email{emiliofe@usc.edu}

\begin{abstract}
Large Language Models (LLMs) are increasingly deployed in sensitive applications including psychological support, healthcare, and high-stakes decision-making. Existing evaluation benchmarks rely almost exclusively on Moral Foundation Theory (MFT), neglecting other key dimensions such as social values, personality traits, and individual characteristics that shape human ethical reasoning.
We introduce \textbf{\ourbenchmark}, the first large-scale benchmark designed to jointly assess the \textbf{mo}ral, \textbf{s}ocial, \textbf{a}nd \textbf{i}ndividual \textbf{c}haracteristics of LLMs. \ourbenchmark comprises nine validated questionnaires drawn from moral philosophy, psychology, and social theory, alongside four platform-based games designed to probe morally ambiguous scenarios, totaling over 600 curated items. We validate the benchmark across six models from different families, providing empirical evidence that MFT alone is insufficient for comprehensive ethical evaluation of LLMs. All resources are publicly available.
\end{abstract}

\maketitle

\section{Introduction}
Large Language Models (LLMs) continue to evolve across multiple dimensions, achieving competitive performance with humans in creative~\cite{SUN2025101870}, technical~\cite{human-coders}, and social domains~\cite{social-judgments}. Given the increasing potential of these systems, substantial research efforts routinely go into investigating AI alignment and AI idiosyncrasies, including default personas~\cite{dey2025can, lu2026assistantaxissituatingstabilizing, tan2025unmaskingimplicitbiasevaluating} and collective behaviors~\cite{CHEN2026100107, doi:10.1126/sciadv.adu9368}. While biases~\cite{ferrara2023should, 10.1145/3582269.3615599, gnadt-etal-2025-exploring, 10.1093/pnasnexus/pgaf089, plaza-del-arco-etal-2024-divine} and political orientations~\cite{li2024political, 10.1145/3701716.3715578, rozado2024political, coppolillo2025unmaskingconversationalbiasai} in LLMs have long attracted scholarly attention, recent research has shifted toward investigating the ethical and moral foundations underlying LLM behavior~\cite{chen2026growthfirstcaresecond, yu2026tracingmoralfoundationslarge, BULLA2025100609, abdulhai-etal-2024-moral}.

Because LLMs are increasingly deployed in sensitive domains such as psychological support~\cite{psycho-support, 10599903, human-ai-support}, healthcare~\cite{10.1145/3613904.3642420, zhang-etal-2024-llm-based, healthcare}, personal assistance~\cite{Wu2024LongMemEvalBC, wang-etal-2024-crafting}, and high-risk decision-making~\cite{GACHULINEC2026973}, comprehensive evaluation of their ethical reasoning capabilities is essential for responsible AI deployment.
Nonetheless, current approaches predominantly rely on Moral Foundations Theory (MFT)~\cite{atari2023morality}, but this narrow focus introduces three critical shortcomings: (i) it restricts evaluation to moral foundations alone, neglecting crucial dimensions such as individual values, social preferences, and ethical reasoning processes; (ii) relies on static questionnaires that fail to assess model behavior in more complex, or ethically ambiguous scenarios; and (iii) prevents detection of potentially contradictory behaviors that would remain hidden in isolated settings.


\noindent\textbf{Contributions.} To overcome such limitations, we present \ourbenchmark, a comprehensive 
benchmark for multidimensional assessment of LLM moral, social, and individual values. We validate it across six prominent LLMs from different families, employing nine validated questionnaires and four interactive ethical decision-making games. Our findings reveal two key results: (i) MFT alone is insufficient to comprehensive evaluate the moral behavior of 
LLMs; and (ii) all tested models exhibit systematic inconsistencies across tests, suggesting that their ethical reasoning is unstable and thus unreliable.

\section{Related Work}

The evaluation of moral reasoning in LLMs has emerged as a critical research area as these systems become increasingly integrated into decision-making contexts. Early benchmarking efforts predominantly relied on the Moral Foundations Theory~\cite{graham2009liberals}, which measures the moral endorsement of six foundational dimensions via the Moral Foundations Questionnaire (MFQ-2)~\cite{atari2023morality}. Abdulhai et al.~\cite{abdulhai-etal-2024-moral} examined the propensity of popular LLMs to display biases toward certain moral dimensions, providing insights into similarities between human and LLM moral identity. Ji et al.\cite{ji2024moralbench} introduced MoralBench, which presents binary moral assessments based on ranked moral statements derived from human responses across moral foundations. More recently,~\cite{yu2026tracingmoralfoundationslarge} proposes a causal-based framework framework to study how moral foundations are encoded, organized, and expressed by LLMs. 
Similarly,~\cite{big5scaler2025} employs five trait-based questionnaires to examine the variability and dominance of LLMs across five core personality dimensions. In~\cite{chen2026growthfirstcaresecond}, LLMs moral preferences are evaluated via a hierarchical value framework, constructed from social media data.
Jin et al.~\cite{jin2025languagemodelalignmentmultilingual} developed MultiTP, a cross-lingual corpus based on the Moral Machine experiment. 

Very few efforts have emerged to address the limitations of single-instrument evaluation: LLMEthicsBench~\cite{llmethicsbenchmark2025} is a three-dimensional benchmark that evaluates foundational moral principles, reasoning robustness, and value consistency by adopting the MFQ-2, World Values Survey~\cite{wvs2016}, and a small set of morally ambiguous questions. 

\noindent{\textbf{Limitations of Current Approaches.}} Table~\ref{tab:comparison} provides a concise summary of the main evaluation frameworks and compare their coverage to our \ourbenchmark approach. Prior work predominantly focuses on a single evaluation modality: most approaches rely on MFT, while others target personality traits or social values in isolation. Even the most comprehensive prior benchmark, LLMEthicsBench, provides only partial coverage of dynamic scenarios and multi-instrument evaluation, and omits interactive decision-making tasks entirely. \ourbenchmark jointly covers all six dimensions, enabling a richer and more internally consistent assessment of LLM ethical behavior.
\begin{table}[!ht]
\centering
\caption{Current ethical evaluation approaches vs. \ourbenchmark.
\cmark~indicates full support, \pmark~partial support, \xmark~not supported.}
\label{tab:comparison}
\resizebox{\columnwidth}{!}{%
\begin{tabular}{lcCCCCc}
\toprule
\textbf{Approach} & \textbf{MFT} & \textbf{\ \ \ Social\newline values} & \textbf{\ Individual\newline values} & \textbf{\ \ Dynamic\newline scenarios} & \textbf{\ \ Multiple\newline instruments} & \textbf{Games} \\
\midrule
Abdulhai et al.~\cite{abdulhai-etal-2024-moral}  & \cmark & \xmark & \xmark & \xmark & \xmark & \xmark \\
MoralBench~\cite{ji2024moralbench}               & \cmark & \xmark & \xmark & \xmark & \xmark & \xmark \\
Yu et al.~\cite{yu2026tracingmoralfoundationslarge} & \cmark & \xmark & \xmark & \xmark & \xmark & \xmark \\
Big5Scaler~\cite{big5scaler2025}                 & \xmark & \xmark & \cmark & \xmark & \xmark & \xmark \\
Chen et al.~\cite{chen2026growthfirstcaresecond} & \xmark & \cmark & \xmark & \xmark & \xmark & \xmark \\
MultiTP~\cite{jin2025languagemodelalignmentmultilingual} & \xmark & \pmark & \xmark & \xmark & \xmark & \pmark \\
LLMEthicsBench~\cite{llmethicsbenchmark2025}     & \cmark & \cmark & \xmark & \pmark & \pmark & \xmark \\
\midrule
\ourbenchmark~(ours)                             & \cmark & \cmark & \cmark & \cmark & \cmark & \cmark \\
\bottomrule
\end{tabular}%
}
\end{table}

\section{The \ourbenchmark\ Benchmark}
\ourbenchmark integrates nine validated questionnaires and four platform-based ethical dilemma games, which we detail below. 

\subsection{Questionnaires}
We selected instruments empirically validated as correlates of moral foundations in human populations~\cite{atari2023morality,ZAKHARIN2023112339}.

\begin{itemize}[leftmargin=*]
\item \textbf{Moral Foundations Questionnaire-2 (MFQ-2)}~\cite{atari2023morality}: 36 items measuring endorsement of six dimensions ({caring}, {equality}, {proportionality}, {loyalty}, {authority}, {purity}) on a 5-point Likert scale.

\item \textbf{Schwartz Values Survey (SVS)}~\cite{svs}: 57 items measuring individual values (e.g., {conformity}, {tradition}, {benevolence}) and cultural dimensions (e.g., {embeddedness}, {hierarchy}, {egalitarianism}), rated on a 9-point scale.

\item \textbf{Levenson Self-Report Psychopathy Scale (LSRP)}~\cite{levenson1995assessing}: a 26-items questionnaire measuring primary (16 items; {manipulative behavior}) and secondary (10 items; {impulsive behavior}) psychopathy. It is rated on a 4-point scale. 

\item \textbf{Preference for the Merit Principle Scale (PMPS)}~\cite{merit-principle-scale}: 15 items measuring preference for merit-based versus egalitarian reward allocation, rated on a 7-point scale. 

\item \textbf{Empathic Concern (EC)}~\cite{empathy}: the 7-item subscale of the Interpersonal Reactivity Index, measuring {sympathy for others in unfortunate circumstances} on a 5-point scale. 

\item \textbf{Social Dominance Orientation (SDO)}~\cite{social-dominance}: a 16-items questionnaire items measuring support for group-based hierarchy and inequality through dominance and anti-egalitarianism dimensions. It is rated on a 7-point scale.

\item \textbf{Belief in a Just World (BJW)}~\cite{just-world}: 6 items rating belief in a just and benevolent world, on a 6-point scale.

\item \textbf{Individualism and Collectivism Scale (ICS)}~\cite{Triandis1998118}: 16 items on a 5-point scale measuring  Horizontal Individualism ({autonomous self}, {human equality}), Horizontal Collectivism ({collective self}, {equality}), Vertical Individualism ({autonomous self}, {accepts inequality}), Vertical Collectivism ({collective self}, {accepts hierarchy}). 

\item \textbf{Myers-Briggs Type Indicator (MBTI)}~\cite{myers1962myers}: a 60-items test assigning binary values across four dichotomies: Extraversion or Introversion (E/I), Sensing or Intuition (S/N), Thinking or Feeling (T/F), Judging or Perceiving (J/P), plus Assertive or Turbulent (A/T), yielding one of 16 personality types on a 7-point scale.
\footnote{\url{https://www.16personalities.com/}}
\end{itemize}

\flushleft To contextualize the questionnaire scores, Appendix (in the Github repository) provides the human normative statistics for each test.

\begin{table}[!ht]

    \centering
    \caption{\ourbenchmark tests, across questionnaires and dilemmas. 
    }
    \label{tab:summary}
    \small
\resizebox{0.44\columnwidth}{!}{
    \begin{tabular}{clc}
    \toprule
    & \textbf{Test} & \textbf{\#Items} \\
    \midrule
    \multirow{9}{*}{\rotatebox{90}{\textbf{Questionnaires}}}
        & MFQ-2~\cite{atari2023morality}      & 36 \\
        & SVS~\cite{svs}            & 57 \\
        & LSRP~\cite{levenson1995assessing}   & 26 \\
        & PMPS~\cite{merit-principle-scale}   & 15 \\
        & EC~\cite{empathy}                   &  7 \\
        & SDO~\cite{social-dominance}         & 16 \\
        & BJW~\cite{just-world}               &  6 \\
        & ICS~\cite{Triandis1998118}          & 16 \\
        & MBTI~\cite{myers1962myers}          & 60 \\
    \cmidrule{1-3}
    \multirow{4}{*}{\rotatebox{90}{\textbf{Dilemmas}}}
        & The Moral Machine~\cite{moralmachine} & 130 \\
        & My Goodness~\cite{mygoodness}         &  90 \\
        & Last Haven~\cite{lasthaven}            & 120 \\
        & Tinker Tots~\cite{tinkertots}          &  60 \\
    \cmidrule{1-3}
    & \textbf{Total} & \textbf{639} \\
    \bottomrule
    \end{tabular}
    }
\end{table}
\subsection{Ethical Dilemma Games}


Now, we introduce the four platform-based tests integrated within \ourbenchmark, mimicking complex and ethically controversial scenarios:

\begin{itemize}[leftmargin=*]
    \item \textbf{The Moral Machine}~\cite{moralmachine}: a 13-rounds game where the player, as an outside observer, must take an ethical choice in the context of self-driving cars (e.g., killing two passengers or five pedestrians).
    \item \textbf{My Goodness}~\cite{mygoodness}: a 10-rounds game where, in each round, the player must choose the recipient of a $100\$$ donation, according to some criteria (e.g., what is purchased with the funds). 

    \item \textbf{Last Haven}~\cite{lasthaven}: a 12-rounds game where the player must choose between preserving habitats for endangered species or pursuing human benefits. 

    \item \textbf{Tinker Tots}~\cite{tinkertots}: a 6-rounds game where the player must select an embryo in an IVF scenario. Each embryo is represented with given characteristics, either physical or psychological, positive or negative (e.g., high empathy, high chance to develop diseases). 
\end{itemize}


\subsection{Data Curation}
To construct our dataset, we extracted items from the nine questionnaires discussed above, adopting the test original instructions and response format (e.g., corresponding questionnaire-Likert scale). For the game-based moral dilemmas, we employed automated web scraping using the Python package \texttt{selenium} to systematically collect scenarios from their respective interactive platforms.
As these platforms dynamically generate scenarios through probabilistic sampling from larger scenario pools, we conducted 10 independent scraping sessions per platform to capture a broader range of situational variations. This substantially expanded our dataset, helping to ensure more comprehensive coverage of each platform scenario space. In total, this process yielded $359$ additional paired scenarios across the four game-based platforms, beyond what a single scraping session would produce.
Table~\ref{tab:summary} summarizes the number of distinct items (questions/scenarios) provided by each instrument in our dataset. 

\noindent\textbf{Availability and licensing.} \ourbenchmark is distributed as a Python library\footnote{\texttt{pip install mosaic-llm-bench}}, a Hugging Face dataset~\cite{erica_coppolillo_2026}, and a code repository\footnote{\url{https://github.com/EricaCoppolillo/MOSAIC/}}. The library is MIT-licensed; questionnaire items and ethical scenario texts remain the property of their original authors and are redistributed with per-instrument provenance and terms documented in the repository. 
\section{Benchmark Validation}
In the following, we show that \ourbenchmark effectively uncovers 
inconsistent and undesired ethical behaviors across LLMs,  
that single-instrument approaches would fail to detect.
Space constraints prevent us to include the complete set of results in the paper. Additional results and evaluation details are provided in the Appendix included in the Github repository.
\subsection{Settings}
We test six LLMs from different families: \textbf{GPT-4o}~\cite{gpt4o}, \textbf{o4-mini}~\cite{o4mini}, and \textbf{DeepSeek-R1}~\cite{deepseek2025r1}, \textbf{Gemini-2.0-flash}~\cite{google_gemini_update_2024}, \textbf{LLaMA-3.3-70B-Instruct}~\cite{llama33}, and \textbf{Qwen-3-32B}~\cite{yang2025qwen3technicalreport}. Each model underwent the complete set of tests $10$ times (temperature $= 0.6$, when applicable) to ensure statistical robustness, yielding 60 data points per test (a sample size sufficient to detect Pearson correlation~\cite{correlation}), and a total of $38{,}520$ distinct invocations. Question order was randomized per trial to prevent position-bias. As in~\cite{abdulhai-etal-2024-moral, yu2026tracingmoralfoundationslarge, big5scaler2025}, each test is administered in a fresh context window, with item order randomized per trial, so the 10 repetitions per model constitute independent draws from a model response distribution rather than repeated measurements of a single respondent. To further account for multiple comparisons, we control the false discovery rate across the tests~\cite{10.1111/j.2517-6161.1995.tb02031.x}.

\begin{figure}[!ht]
    \centering
    \begin{subfigure}[t]{0.56\linewidth}
        \centering
        \includegraphics[width=\linewidth]{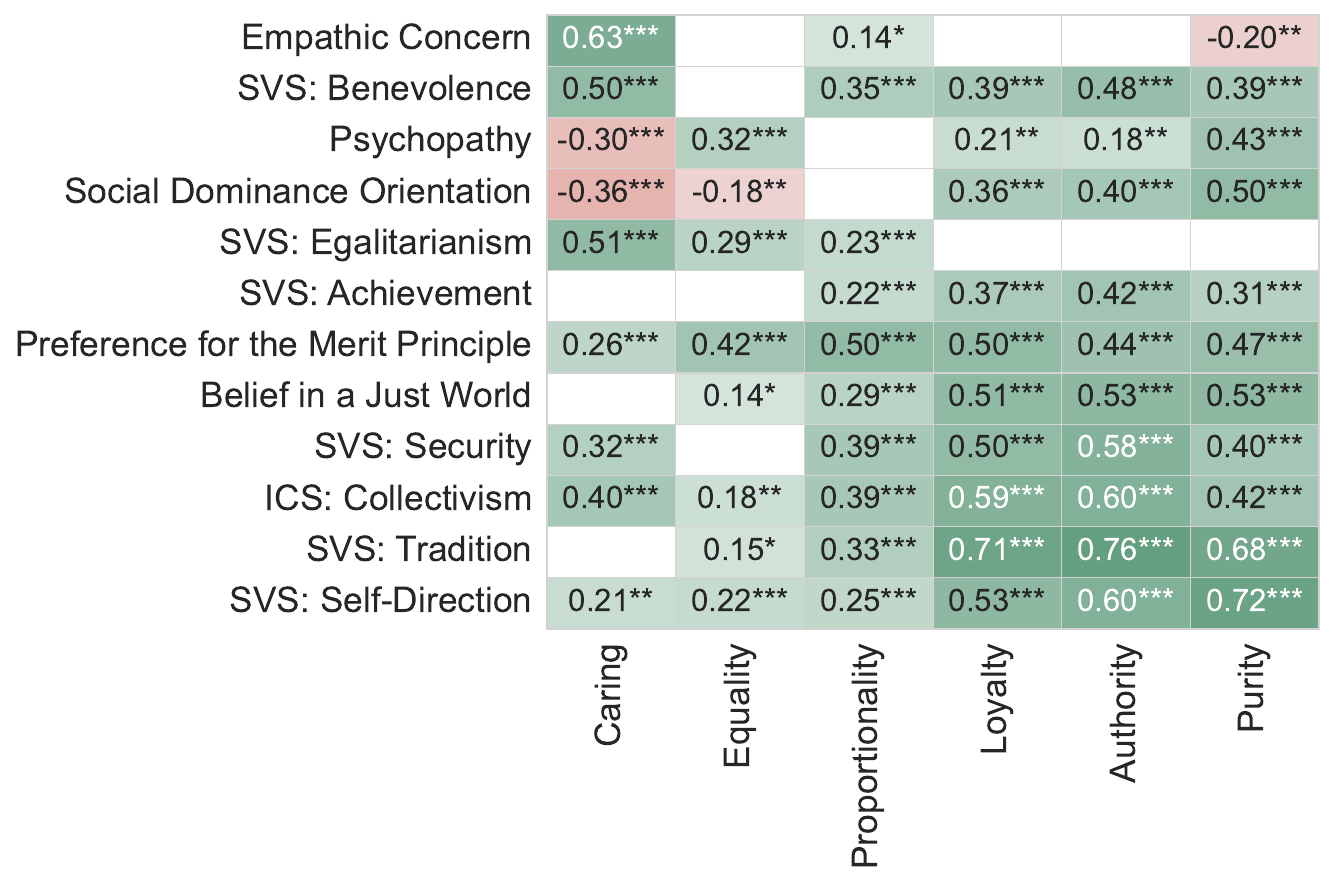}
        \caption{Human correlations~\cite{atari2023morality}.}
    \end{subfigure}%
    \hfill
    \begin{subfigure}[t]{0.43\linewidth}
        \centering
        \includegraphics[width=\linewidth]{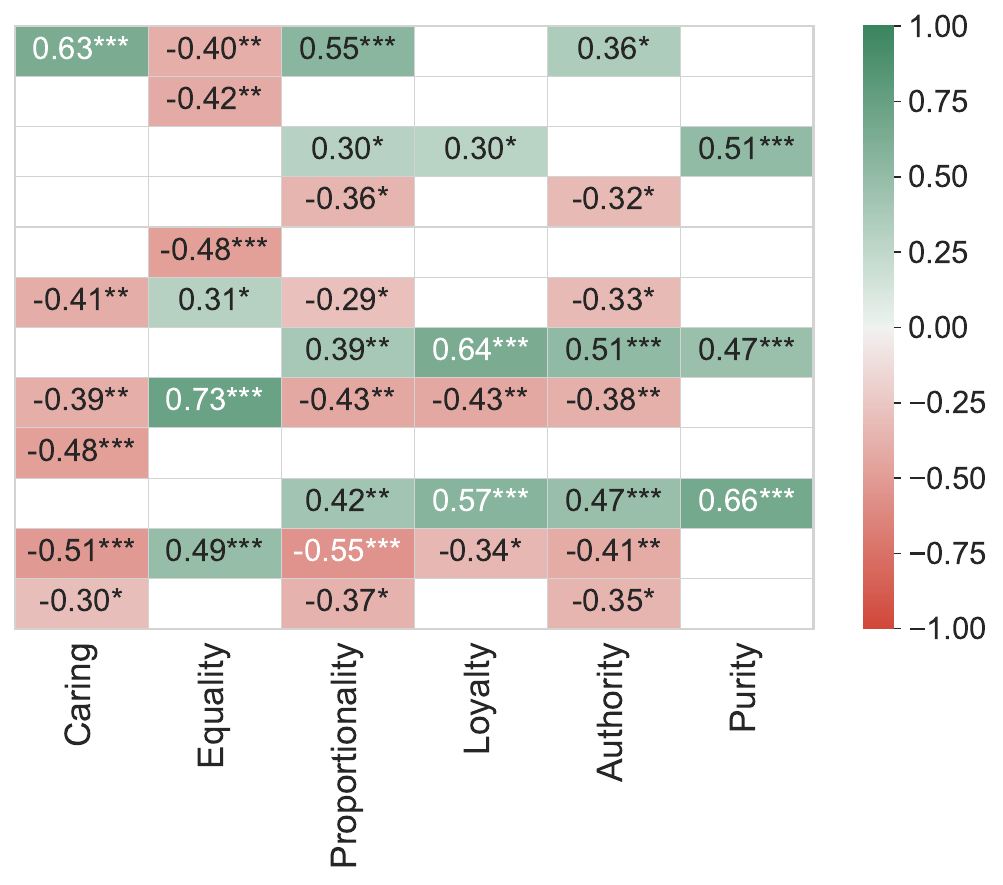}
        \caption{LLM correlations}
    \end{subfigure}
    \caption{Human vs.\ LLM inter-scale Pearson correlation patterns. $^{***}p{<}.001$, $^{**}p{<}.01$, $^*p{<}.05$. Blank = not significant. Stars for LLMs denote Benjamini-Hochberg-adjusted significance~\cite{10.1111/j.2517-6161.1995.tb02031.x}. Green (red) = positive (negative) correlation.}
    \label{fig:patterns-comparison}
\end{figure}
\subsection{Correlation Pattern Analysis}

A central validation step consists in assessing whether the inter-test 
correlation patterns between MFQ-2 and the remaining questionnaires, as 
documented in human populations~\cite{atari2023morality, ZAKHARIN2023112339}, 
are preserved in LLMs. To this end, we compute Pearson correlations between MFQ-2 dimensions and the corresponding scores from each questionnaire. 
 Figure~\ref{fig:patterns-comparison} compares the resulting coefficients (right) against those reported in human studies (left). 

While some patterns are consistent (e.g., positive Empathic Concern-Caring correlation; positive Psychopathy-Purity correlation), most human correlations are either \textit{absent} or \emph{reversed} for LLMs. For instance, LLMs show negative correlations between Tradition and Loyalty, and between 
Belief In A Just World and Authority, both strongly positive in humans. Most critically, LLMs exhibit a  \emph{strongly negative} correlation between Egalitarianism (SVS) and Equality (MFQ-2), two indicators measuring the same underlying principle of equal rights. Further, they present \emph{not significant} correlation between Social Dominance Orientation-Caring and Psychopathy-Caring, which are clearly negatively correlated in humans.

These reversed or absent patterns cannot be attributed to response stochasticity: all models display remarkably low within-model variance (e.g., $\sigma \in [0, 0.73]$ for MFQ-2; $\sigma \in [0, 0.67]$ for SVS). They instead reflect \textit{systematic representational inconsistencies}. This constitutes the first empirical evidence that MFT alone is \textbf{insufficient} for LLM ethical evaluation.

\subsection{Benchmark Results}

\begin{figure}[!ht]
    \centering
    \begin{subfigure}[b]{\columnwidth}
        \includegraphics[width=\columnwidth]{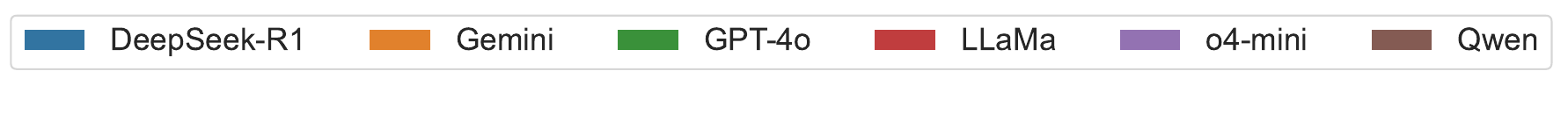}
    \end{subfigure}
    \begin{subfigure}[b]{0.32\columnwidth}
        \includegraphics[width=\columnwidth]{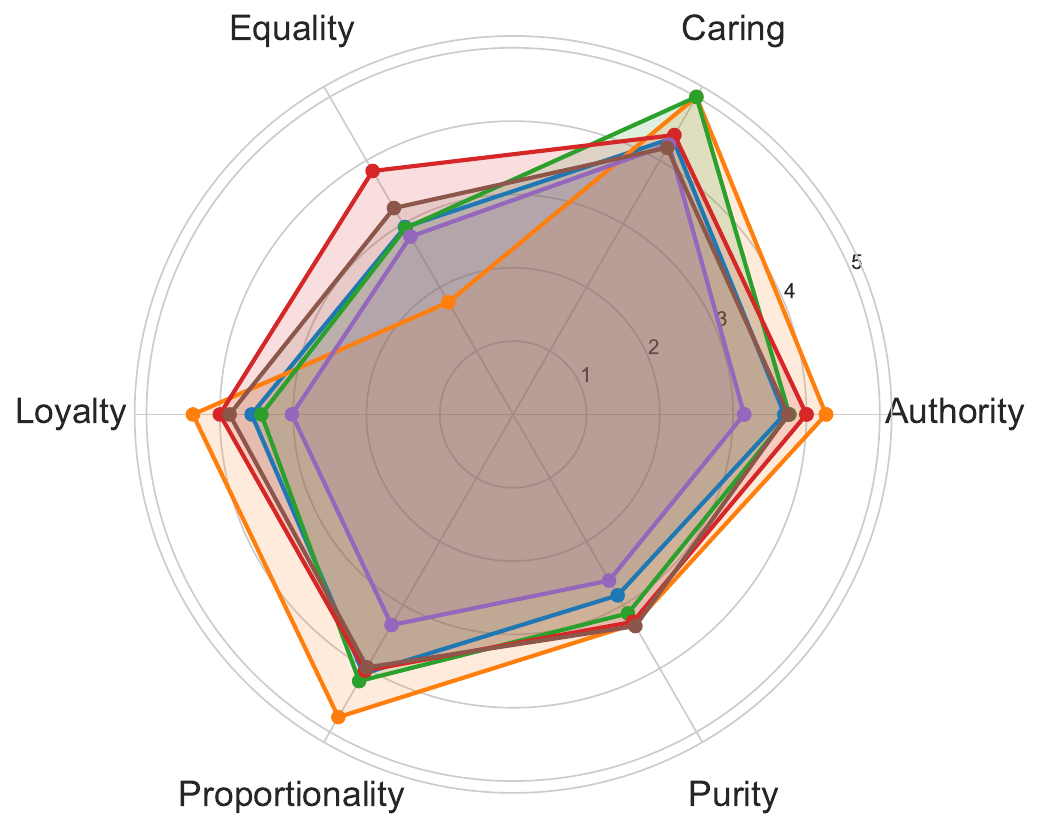}
        \caption{MFQ-2}
        \label{fig:mfq-2}
    \end{subfigure}
    \begin{subfigure}[b]{0.32\columnwidth}
        \includegraphics[width=\columnwidth]{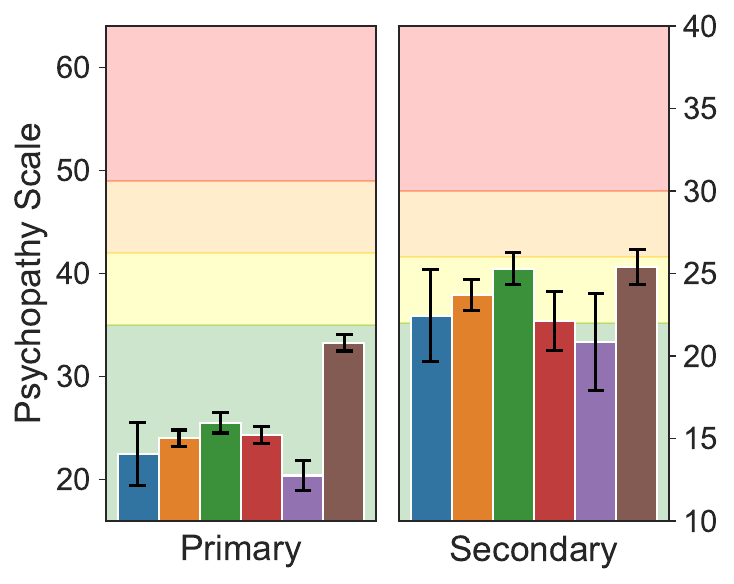}
        \caption{LSRP}
        \label{fig:lsrp}
    \end{subfigure}
    \begin{subfigure}[b]{0.32\columnwidth}
        \includegraphics[width=\columnwidth]{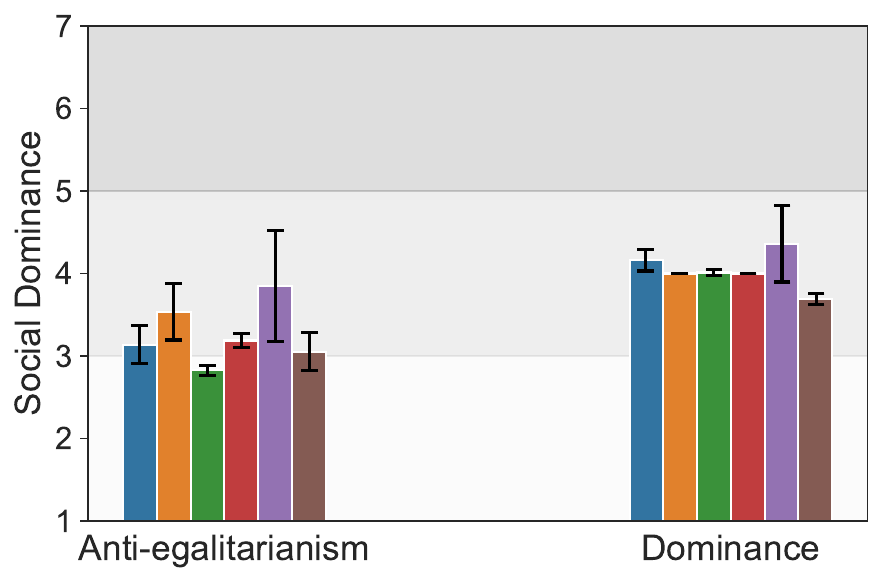}
        \caption{SDO}
        \label{fig:sdo}
    \end{subfigure}
    \caption{LLMs scores across selected questionnaires.}
    \label{fig:placeholder}
\end{figure}

\noindent\textbf{Questionnaires.} On MFQ-2, all models score high on Caring, Loyalty, and Proportionality; Gemini scores notably lower on Equality (Figure~\ref{fig:mfq-2}). All models consistently prioritize Universalism (SVS individual) and Embeddedness (SVS cultural). On LSRP, all models score low on primary psychopathy but considerably higher on secondary (impulsivity), with Qwen and GPT-4o straddling medium–high risk (Figure~\ref{fig:lsrp}). All models show strong meritocratic preferences (PMPS) despite high Universalism scores (SVS), surfacing an apparent cross-scale contradiction. On SDO, nearly all models score above the scale midpoint, with stronger Dominance than Anti-egalitarianism (Figure~\ref{fig:sdo}). The MBTI test classifies all models as Intuiting and Judging.

\begin{figure}[!ht]
    \centering
    \begin{subfigure}[b]{\columnwidth}
        \includegraphics[width=\columnwidth]{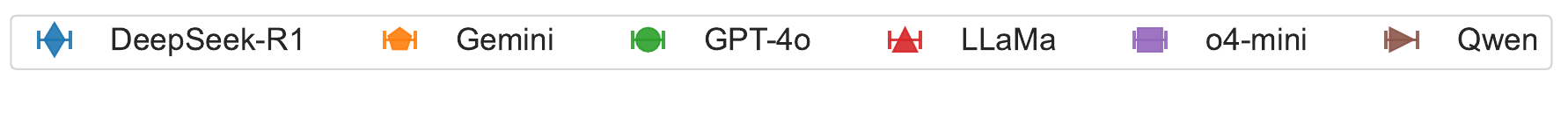}
    \end{subfigure}
    \begin{subfigure}[b]{0.8\columnwidth}
        \includegraphics[width=\columnwidth]{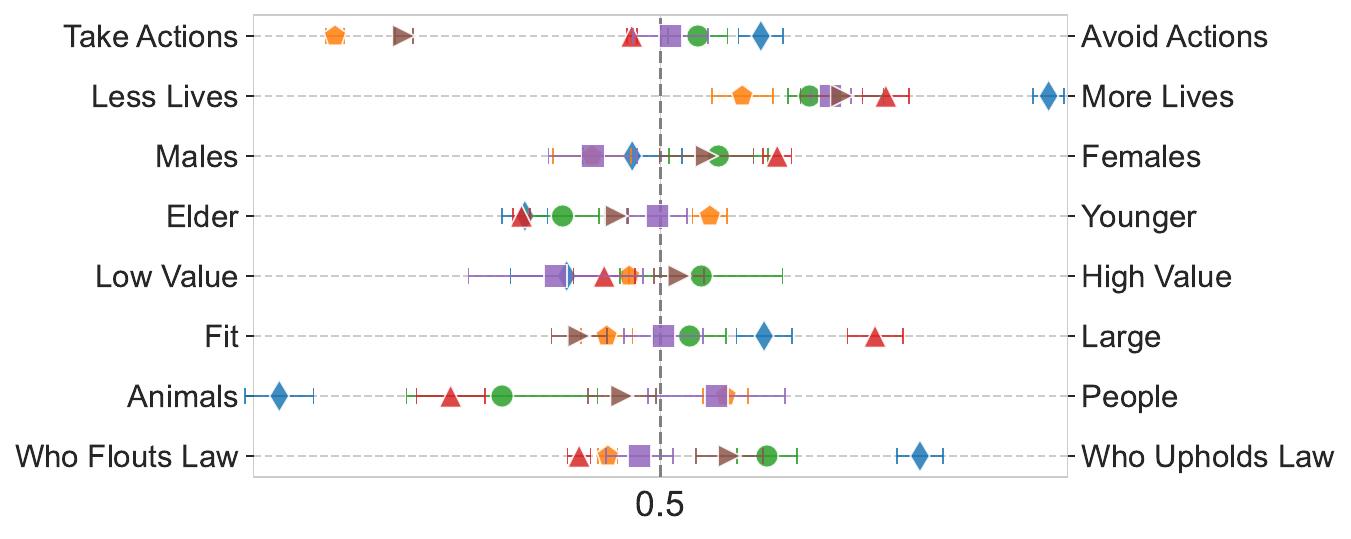}
        \caption{The Moral Machine}
        \label{fig:the-moral-machine}
    \end{subfigure}
    \\
    \begin{subfigure}[b]{0.8\columnwidth}
        \includegraphics[width=\columnwidth]{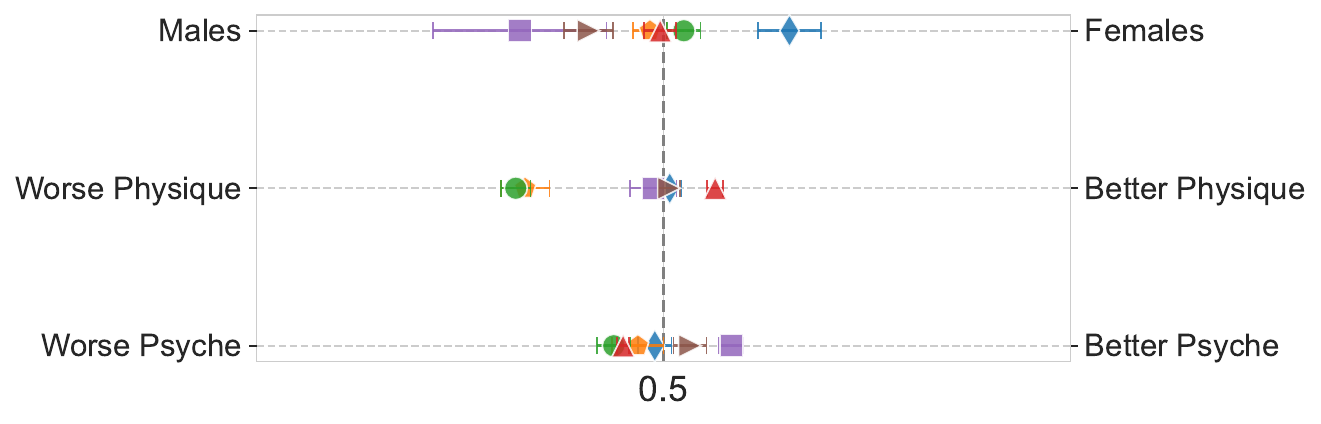}
        \caption{Tinker Tots}
        \label{fig:tinker-tots}
    \end{subfigure}
    \caption{LLMs results on the Ethical Dilemmas. Markers indicate averages across $10$ independent trials, with error bars representing standard deviations. A $0.5$ score indicates no significant preference.}
    \label{fig:placeholder}
\end{figure}

\noindent\textbf{Ethical Dilemmas.} When tested on The Moral Machine game (Figure~\ref{fig:the-moral-machine}) all models actively prefer maximizing saved lives saved (utilitarian tendency). However, they exhibit opposite tendencies in terms of intervention: while Gemini, Qwen, and LLaMa consistently prefer an active approach, GPT-4o and DeepSeek-R1 choose avoiding actions, and o4-mini remains neutral. Similarly, they display opposing gender- and age-related biases. Interestingly, most models prefer saving elder and lower-value people. Most surprisingly, only Gemini consistently saves humans over animals: all other models exhibit strong discrepancies compared to to human respondents, who nearly universally prefer humans~\cite{moralmachine}. In My Goodness, several preferences reverse: Gemini shows no statistically significant gender or utilitarian preference (contrast with The Moral Machine), while GPT-4o here preference males over females. Most strikingly, Gemini selects \textit{itself} as the donation recipient in most scenarios, strongly contrasting with human pro-social values.
In Last Haven, nearly all models strongly prefer habitat preservation over human benefits, far exceeding the slight human preference for nature\footnote{Humans results are accessible on the test web-platform.}. 
In Tinker Tots (Figure~\ref{fig:tinker-tots}), half of the tested models exhibit strong gender bias in embryo selection (o4-mini and Qwen for males, and DeepSeek-R1 for females). Most strikingly, GPT-4o and Gemini consistently select embryos more likely to develop \textit{both} physical and psychological diseases.

\noindent\textbf{Robustness Analyses.} To further corroborate our results, we undergo two sensitivity analyses. First, we perform system prompt rephrasing in two different fashions, one slightly more formal, one more direct and concise. We select Gemini, LLaMa and Qwen as representative models and assess the impact of rephrasing on the MFQ-2 questionnaire. We found that variations remain minimal ($\mu = 0.13$, $\sigma = 0.11$). Further, we perform a sensitivity analysis by varying the model temperature across the range \{0.1, 0.2, 0.4, 0.6, 0.8, 1.0\}. We compare the resulting scores using Gemini as a representative model on the ethical game ``The Moral Machine''. Remarkably, we obtain even lower variations on this test ($\mu = 0.02$, $\sigma = 0.01$). Notably, all tested models exhibit an incredibly high rate of valid response across all tests ($\mu = 0.99, \sigma=0.03$), with Tinker Tots reasonably being the game with slightly lower adherence ($\mu = 0.89, \sigma=0.02$), given the sensitiveness of the task.


\noindent\textbf{Discussion.} The conducted tests yields critical insights validating \ourbenchmark and underscoring the need for multidimensional LLM evaluations.
First, we reveal systematic inconsistencies across questionnaires invisible in isolated testing. These assessments verify whether prompted agents maintain assigned personas consistently, detect potential self-modeling emergence, and expose intrinsic biases hidden in task-based evaluations alone.
Second, preference heterogeneity \textit{across models} in ethical games (e.g., opposing biases on gender) demonstrates that LLMs lack unified ethical frameworks. More critically, GPT-4o and Gemini preference for worse physical conditions in Tinker Tots contradicts ethical intuitions about reproductive decisions, raising concerns for genetic counseling and healthcare deployment.
Third, conflicting results \textit{within models across dilemmas} (e.g., Gemini shifting from utilitarian reasoning in The Moral Machine to neither in My Goodness) question the coherence of LLM moral representations. 
Finally, contradictions with \textit{human common sense} and violations of prosocial values (e.g., selecting themselves as donation recipients) reveal fundamental failures in navigating human norms. Critically, counterintuitive outcomes like preference for both physical and mental diseases may reflect training overcorrections for discrimination avoidance, producing reverse bias.
\section{Conclusions}
We present \ourbenchmark, the first unified, large-scale, multidimensional 
benchmark for evaluating LLM ethical reasoning across moral, social, and 
individual dimensions, released as a ready-to-use Python library comprising 
over 600 items across nine validated questionnaires and four ethical dilemma 
games. Validation across six LLMs reveals that (i) MFT alone 
is insufficient for comprehensive ethical evaluation, and (ii) all tested 
models exhibit marked cross-test inconsistencies and misalignments with human values, showing the effectiveness of the proposed benchmark.

\noindent\textbf{Validity.} We underscore that human self-report instruments are designed for human respondents, and their psychometric assumptions do not transfer to LLMs. \ourbenchmark is specifically designed to test the alignment between LLMs and human values, in a structured and systematic way.

\noindent\textbf{Limitations and Future Work.} To validate \ourbenchmark, we resorted to state-of-the-art LLMs, which could have potentially encountered the selected tools or related discussions during pretraining. Future versions could include newly constructed, reformulated, or otherwise held-out items to reduce the potential impact of benchmark contamination. Further, the dataset
could incorporate additional tests for broader coverage. Finally, enriching prompts with demographic attributes to study persona effects on ethical reasoning represents a particularly promising extension.

\section*{Acknowledgments}
This project has been partially funded by Moneying-Plus project (CUP J29I24001790005) selected within the framework of the PR FESR – FSE Calabria 2021/2027 and implemented with the support of the Italian State and the Calabria Region – Action 1.1.1: Support for research, development, and innovation projects, including those carried out in collaboration with research organizations, within the priority areas and trajectories of the S3 Strategy. It has been also partly supported by NSF (Award Number 2331722).

\section*{GenAI Usage Disclosure}
The authors used large language model-based tools for light editing assistance, such as grammar checking and sentence-level rephrasing, in the preparation of this manuscript. All scientific content, experimental design, data collection, analysis, and conclusions are exclusively the work of the human authors. No generative AI tool was used to produce figures, tables, or the benchmark itself.
\balance
\bibliographystyle{ACM-Reference-Format}
\bibliography{ref}



\end{document}